\documentclass[12pt,preprint]{aastex}
\usepackage{graphicx,amssymb}%,amsmath,amsfonts}  

\begin{document}

\shorttitle{Effects of black hole's grav. field on the luminosity 
of a star}
\shortauthors{Gomboc \& \v Cade\v z}

\title{Effects of black hole's gravitational field on the luminosity 
of a star during close encounter}
\author{Andreja Gomboc$^{1,2}$ \and Andrej \v Cade\v z$^{1}$}
\affil{Faculty of Mathematics and Physics, University in Ljubljana, 
Jadranska 19, 1000 Ljubljana, Slovenia \and Astrophysics Research Institute, 
Liverpool John Moores University, 12 Quays House, Egerton Wharf, Birkenhead, 
CH 41 1LD, UK}
\email{andreja.gomboc@fmf.uni-lj.si, andrej.cadez@fmf.uni-lj.si}

%\date{Received   / Accepted }

\begin{abstract}
To complement hydrodynamic studies
of the tidal disruption of the star by a massive black hole, we present the study of stellar 
luminosity and its variations, produced by the strong gravitational field of the black hole 
during a close encounter. 
By simulating the relativistically 
moving star and its emitted light and taking into account general relativistic effects 
on particle and light trajectories, our results show that the black hole's gravity alone
induces apparent stellar 
luminosity variations on typical timescales of a few r$_g$/c (=5 sec $\rm{{{m_{bh}}\over {10^6 M_\odot}}}$) to a
few 100 r$_g$/c ($\sim $ 10 min $\rm{{{m_{bh}}\over {10^6 M_\odot}}}$), where r$_g$=Gm$_{bh}$/c$^2$. 
We discern different cases 
with respect to the strength of tidal interaction and focus on two: 
a) a star encountering a giant black hole traces space-time almost as a point particle, 
so that the apparent luminosity variations  are dominated by clearly recognizable general
relativistic effects and b) 
in a close encounter of a star with a black hole of similar size the stellar debris is spread 
about the black hole by processes where hydrodynamics plays an important role.
We discuss limitations and results of our approach. 

\end{abstract}

\section{Introduction}
Motivation for our work comes from the presence of massive black 
holes in galactic nuclei and from the possibility that such black holes 
accrete material from their surroundings. It was estimated (Gurzadyan 
\& Ozernoy 1981; Rees 1990; Magorrian \& Tremaine 1999; Syer \& Ulmer 1999) that central
black holes may capture stars from inner galactic regions at the rate 
from $\rm{10^{-3}}$ to $\rm{10^{-7}}$ per galaxy per year. 
Such events would be particularly interesting in Galactic centre, where the observed X-ray flare 
(Baganoff et al. 2001) and measured motion of stars, down to only 17 light hours from 
the centre (Sch\"odel et al. 2002), provide strong evidence that the central 
concentration of about 3$\times$10$^{\rm{6}}$~M$_\odot$ is indeed a black hole.
In recent years UV and X-ray flares have been observed in the nuclei of
NGC 4552, NGC 5905, RX J1242.6-1119, RX J1624.9+7554 and others, for which it
was concluded that tidal disruption of a star by a massive black hole provides 
the best explanation (Renzini et al. (1995), Komossa \& Bade (1999), Grupe et al. (1999),
Gezari et al. (2003)).

The interaction of a star with a black hole has been studied previously
by other authors (Rees 1988; Carter \& Luminet 1985; Luminet \& Marck 1985)
with a number of detailed hydrodynamics simulations
(Laguna et al. 1993, Khokhlov et al. 1993a, 1993b, Kochanek 1994, Fulbright et al. 1995, Marck et al. 1996, 
Diener et al. 1997, Loeb \& Ulmer 1997, Ayal et al. 2000, Ivanov \& Novikov 2001, 
Ivanov et al. 2003) 
with emphasis on stellar structure during the encounter with the black hole and longterm 
evolution of stellar debris.
Nevertheless, none of these studied the luminosity variations occurring to the
star in the vicinity of the black hole. In order to be complete, such study should include
stellar hydrodynamics in full general relativity, modeling of radiation
processes in the disrupted star and relativistic effects on the emmitted light.
Due to the complexity of the subject, we do not attempt to study all these effects in full
here, but we wish to complement hydrodynamic studies by previously mentioned authors. 
Therefore we limit our attention in this paper to effects on star's luminosity induced
solely by the gravity of the black hole, as we expect that relativistic 
effects alone might produce interesting luminosity phenomena. 
We simulate the disruption and the appearance of the star during close encounter as it 
would be seen by a distant observer and
make a comparison of some results in our model
with those obtained by hydrodynamic simulations.

\vspace{0.5cm}
The model of the star used in our simulations depends on the expected 
strength of the tidal interactions between the star and black hole.
Tidal disruption of the star with mass ${\rm M_*}$ and radius $\rm{R_*}$ occurs 
only if the star approaches the black hole to within its Roche radius:
\begin{equation}
r_R=\Biggl({{m_{bh}}\over {M_*}}\Biggr)^{1\over 3}~ R_* \label{Roche}
\end{equation}
which, expressed in units of the black hole's gravitational radius $r_g=Gm_{bh}/c^2$, reads:
\begin{equation}
\mathcal{R}_R= {{r_R}\over 2r_g} = 25 \Bigl( {{\rho_\odot}\over {\rho_*}}\Bigr)^{1\over 3}~\Bigl({{10^6 M_\odot}\over {m_{bh}}}\Bigr)^{2\over 3}, \label{RocheM}
\end{equation}
where $\rho_\odot $ and $\rho_*$ are the average densities of the Sun and the star. 
It is convenient to introduce the dimensionless Roche radius penetration factor $\beta = r_R/r_p$, where $r_p$ is the periastron distance of the star with respect to the black hole. The Roche penetration factor of a black hole grazing orbit is obviously: $\beta_{gr} = r_R/(2 r_g+\rm {R_*})~=~\mathcal{R}_R/(1+{\rm {R_*}\over {2 r_g}})$. It is shown in the Appendix that this penetration factor crucially determines the strength of tidal interaction, i.e. the amount of work the tidal forces do on the star. We show (eq.{\ref{tidalref}}) that tidal work can be approximated by:
\begin{equation}
W_{tide}~\sim~ G{m_{bh}M_*R_*^2\over r_p^3}~\varepsilon ^2(\beta)~=~ M_* c^2 {r_g \over r_{p}}{R_*^2\over r_p^2}~\varepsilon^2 (\beta)
,
\label{tide}
\end{equation}
where $\varepsilon (\beta)$ can be thought of as an effective eccentricity of the star at the periastron. If the Roche radius penetration factor is large, $\varepsilon $ may grow to values of order $1$, bringing $W_{tide}$ to values comparable to a sizeable fraction of $M_{*} c^2 $. Thus, the tidal interaction becomes overwhelmingly strong for large $\beta$. Such an extreme scenario occurs for grazing interactions only if the size of 
the star is comparable to that of the black hole (see Appendix). We clasify grazing tidal interactions as follows:
\begin{itemize}
\item
{\bf $m_{bh}/M_*\ll 1$}: the Roche radius is smaller than the radius of the star, it follows that the Roche penetration factor is less than 1. As a consequence $\varepsilon^2\ll 1$ and the star as a whole does not suffer large perturbations, even if the black hole pierces the star 
and accretes a small part of its mass along the way. 
\item
{\bf $m_{bh}/ M_*\sim 1$}: the Roche radius more or less equals the radius of the star (eq. \ref{Roche}) and, unless the star is very unusual $R_*\gg r_g$, so that the value of the Roche penetration factor $\beta_{gr} \sim 1$. For such a $\beta_{gr}$, $\varepsilon (\beta=1)\sim 1$ (c.f. Appendix A) and eq. \ref{tide} predicts that the tidal energy is of order $10^{-5}M_*c^2$, which is a typical internal energy of a solar type star. Thus, the tidal energy is just about large enough to completely distort the star; the interaction may 
trigger violent hydrodynamic phenomena, possibly even a supernova. The most important phenomena 
governing the appearance of the star during such an encounter are hydrodynamic in nature, since 
the strong gravity region about the black hole has a much shorter range than is the size of the 
perturbed star. Hydrodynamics governs the appearance of the phenomenon and, therefore, such an event 
does not directly reflect general relativistic effects in strong gravity environment. 
\item
{\bf $m_{bh}/ M_*\sim (c/v_e)^2$}, where $v_e=(2 GM_*/R_*)^{1\over 2}$ is the escape velocity from the star:
the black hole radius is comparable to the size of the star; if the star is not very unusual, its escape velocity is much less than $c$, so that according to eq. \ref{Roche} the Roche radius is much greater than $R_*$ and consequently $\beta_{gr}\gg 1$. In this case the tidal energy exceeds the internal energy by 
several orders of magnitude. A total and complete tidal disruption takes place outside the black hole 
but in the region  sufficiently close to the black hole for relativistic effects to play the mayor 
role in dynamics of the disruption. (Sect. 3). 
\item
{\bf $(c/v_e)^2<m_{bh}/ M_*< (c/v_e)^3$}: 
the black hole radius is larger that that of the star, but still smaller than Roche radius - $\beta_{gr}$ decreases with increasing mass of the black hole. The tidal energy before reaching the horizon is still comparable to the internal energy of the star. The 
release of tidal energy may well be sufficient to produce high energy shocks boosting stellar 
luminosity by many orders of magnitude. Yet, shocks moving with a few Mach are still much slower 
than the near speed of light the star is moving now. The star remains small with respect to 
the black hole along its way to the black hole. Such a stellar capture will thus very closely trace 
relativistic effects in the space-time, as it will be seen almost as a flashing up point particle on 
its way to doom. 
\item
{\bf $m_{bh}/ M_*> (c/v_e)^3$}:
the black hole is very much larger in size than the star ($\rm{m_{bh}>10^8~ M_\odot}$ 
for a Solar type star), the Roche radius lies beyond the black 
hole's horizon, so it follows (\ref{RocheM}) that  
the star is tidally disrupted only after crossing the horizon ($r_R < 2 r_g$). 
Hence,  the point particle approximation for the falling star 
is very good in the whole region outside the black hole. Since there is no agent to heat the star up, it is less likely for such an event to be 
noticed (Sect. 2). 
\end{itemize}

Here we discuss only the last three cases, since we find them interesting as a tool to study the 
strong gravity regions in the universe, as well as
in view of supermassive black holes in galactic nuclei.

\section{Star encountering a $\rm{m_{bh}>10^8~ M_\odot}$ black hole}
We expect that a capture of a star by a giant black hole would most likely occur when a 
star in the cluster surrounding the giant black hole would be perturbed to a 
low angular velocity orbit with respect to the black hole. Therefore, such encounters will, 
most likely occur with the velocity characteristic for parabolic infall. During such an infall 
the star can not be significantly disrupted while outside the horizon, so with respect to a much 
larger black hole it can be treated as a point source of light whose appearance with respect to the 
far observer will be modulated by the Doppler shift, aberration bending and gravitational redshift. 
Two numerical codes were developed to calculate 
the apparent luminosity changes of the source falling in both, 
Schwarzschild type and Kerr type black holes.
During the encounter of the star with a giant black hole the 
star is simulated as a point source emitting monochromatic light of frequency 
$\nu_0$ and intensity $L_0$, both constant in the frame comoving with 
the source. As the source is moving along a parabolic orbit with a given 
orbital angular momentum, we trace light rays from subsequent points 
of the source's trajectory (separated by $\Delta$t=1~$r_g$/c in coordinate time)
to the distant observers and calculate the apparent luminosity with respect to them as a 
function of time. We would like to note,
that these results are directly
applicable also to luminosity and spectrum changes produced by orbiting 
blobs of material in accretion discs around black holes. 

Results for both types of black holes show (Gomboc et al. 1999) 
two characteristic timescales of luminosity changes, 
both determined by the gravity of the black hole. The first one displays the basic  
quasiperiod in luminosity and redshift changes as the star spirals toward the black hole.
The quasiperiod very closely matches the orbital 
period of the source at the innermost stable orbit (50 $r_g$/c for Schwarzschild black hole).
The number of quasiperiods observed depends on the fine tuning of the angular momentum
to the critical value. In Schwarzschild case the critical angular momentum 
is $l/M_* r_g c=\tilde{l}=$4 
and the number of quasiperiods can be approximated as
$N_p=0.5-0.5 Log(4-\tilde{l})$ for 3.9$<\tilde{l}<$4.
The quasiperiods in Kerr case differ for 
prograde and retrograde orbits: for a maximal Kerr black hole (with rotation parameter
a=0.998 $r_g$) and a star on a prograde orbit with angular momentum close to critical 
$\tilde{l}^+\rm{=2(1+\sqrt{1-{a\over {r_g}}})}$, the quasiperiod is $\approx$ 13 $r_g$/c, 
while for a star on a retrograde orbit with $\tilde{l}$ close to critical 
$\tilde{l}^-\rm{=-2(1+\sqrt{1+{a\over {r_g}}})}$, the quasiperiod is $\approx$ 80 $r_g$/c,
both consistent with orbital periods at critical radii.

\begin{figure*}
\centering
\includegraphics[width=17cm]{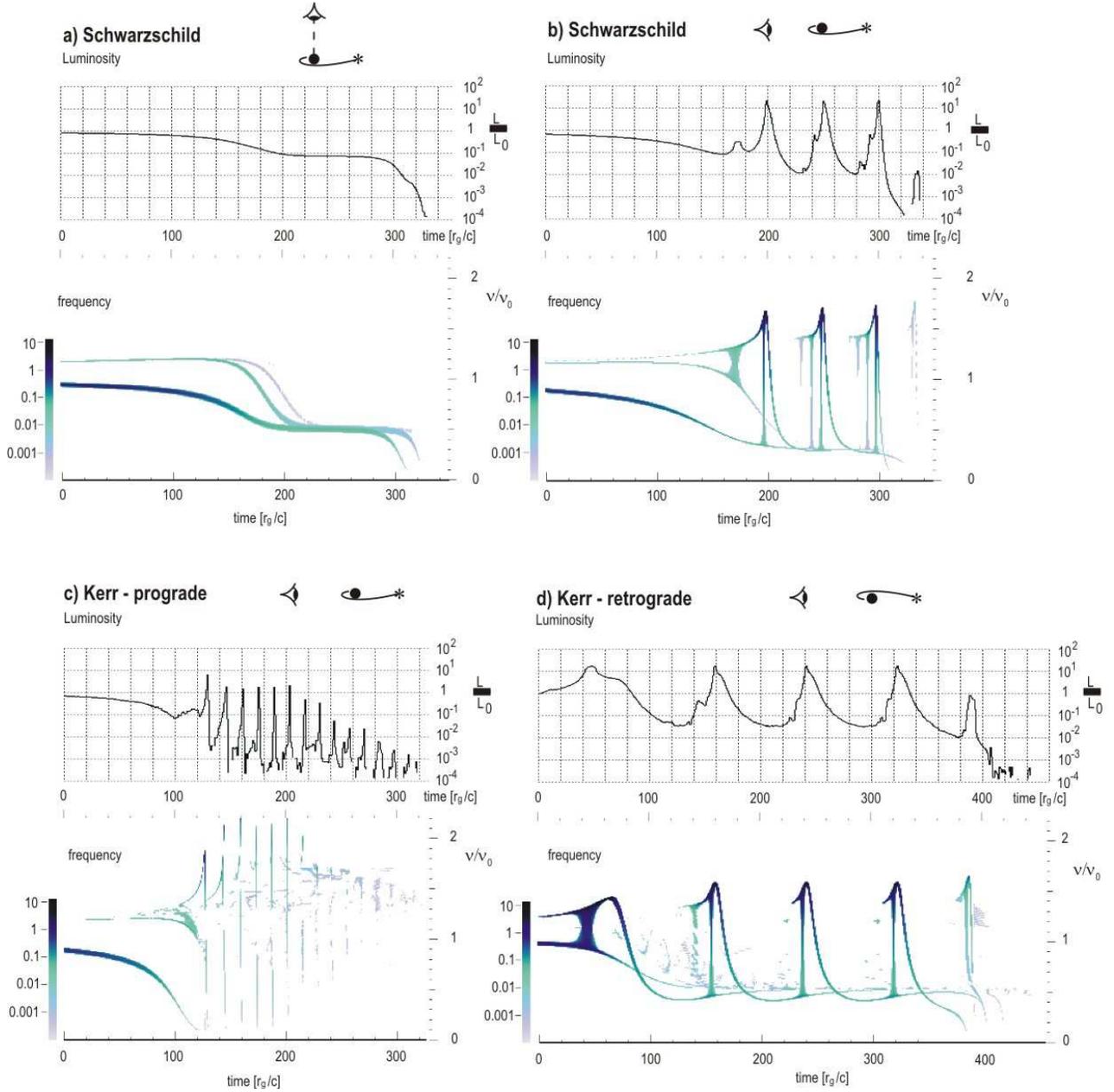}
\caption{Luminosity and frequency shift during the infall of a solar 
type star into a giant black hole $\rm{m_{bh}>10^8~ M_\odot}$. {\bf a)} infall with orbital 
angular momentum $\tilde{l}$ close to critical $\tilde{l}$=4 into the {\bf Schwarzschild} black hole,
as observed perpendicular to the orbital plane; {\bf b)} same event observed in the orbital 
plane; {\bf c)} infall of the star on {\bf prograde} orbit with $\tilde{l}$ close to 
critical $\tilde{l}^+$ into the {\bf Kerr} black hole, as observed in the orbital 
plane; {\bf d)} infall of the star on {\bf retrograde} orbit with $\tilde{l}$ close to 
critical $\tilde{l}^-$ into the {\bf Kerr} 
black hole, as observed in the orbital plane. The color code in 
frequency diagram corresponds to spectrum intensity (in units of the initial intensity of the 
primary image). (Full resolution images available at www.fmf.uni-lj.si/\~~gomboc)}
\label{f1}
\end{figure*}

The second time scale 
is considerably faster (of order $ 1 {r_g}/c$) and belongs to the rate of change of relativistic beaming 
direction with respect to the observer. For the black hole with 
mass $\rm{m_{bh}=10^8~ M_\odot}$, the corresponding timescales are $\sim$ 10 hours 
and $\sim$ 10 min, and for extreme $\rm{m_{bh}=10^{10}~ M_\odot}$, this time 
intervals are $\sim$ months and $\sim$ 10 hours. Since the luminosity and 
spectrum changes are caused by relativistic beaming and gravitational 
lensing, they are most evident to observers in the orbital plane of the 
star. The observers perpendicular to this plane see the source as slowly 
fading and then, as the source approaches the horizon, suddenly disappearing 
on a timescale of the order of $\sim$ 10 $r_g$/c. Comparing results for Schwarzschild 
and Kerr black holes, we find that luminosity curves (Fig. \ref{f1}) 
are qualitatively similar, but timescales generally shorten for Kerr prograde 
orbits and become longer for Kerr retrograde orbits. Results show that within $5^\circ$ of the 
orbital plane one may expect luminosity rise by a factor of a few 10, while the maximum Doppler plus 
redshift factor ($\nu_{obs}/\nu_{0}$) is 1.8 for the Schwarzschild 
case and 2.2 for the maximal Kerr black hole case.

\section{Star encountering a $\rm{m_{bh}\sim 10^5}$ - $\rm{10^6~ M_\odot}$ black hole}

\subsection{Approximations, model and comparison with hydrodynamic results}

The capture of a star by a black hole of comparable size is a phenomenon where black hole's gravity 
plays the dominating role both on propagation of light as well as on propagation of matter 
belonging to the star. This property of the phenomenon is forcefully stressed by the fact that the 
tidal energy is many orders of magnitude larger than its gravitational binding energy and becomes a sizeable fraction of $M_* c^2$ (eq. \ref{tide}).
Therefore, we build our approach on the work of Luminet \& Marck (1985),
who showed that in the vicinity of the black hole "particles of the star undergo a phase of 
approximate free fall in the external gravitational field, since the tidal contribution grows 
much larger than pressure and self gravitating terms". Therefore, 
we use a simple model, whereby the star is considered 
as undisturbed by the black hole (i.e. spherically symmetric), until it reaches 
the Roche radius. After crossing it, the black hole's gravity takes
over and the self gravity and internal pressure are completely switched off.

Further, we neglect hydrodynamic effects. This approximation is 
justified if the proper time elapsed between the Roche radius crossing 
and total disruption is short compared to the dynamic time scale $\tau_d$ of
the star. For an estimate of the two timescales we take
\begin{eqnarray}
\tau_d &=& \Bigl ({{G\varrho_*}\over {3 \pi}}\Bigr)^{-{1\over 2}} \label{tdin}, \\
\tau_R &\sim & {{\sqrt{2} \over {3c}} r_g \cdot \mathcal{R}_R^{3\over 2}} = (6\pi G\varrho_*)^{-{1\over 2}} ,
\end{eqnarray}
where $\tau_R$ is estimated as the proper time elapsed during a 
radial parabolic infall from the Roche radius to the horizon\footnote{Of course, 
$\tau_R$ is defined only for $r_R>2 r_g$, when tidal disruption takes place 
outside the horizon of the black hole. For nonzero angular momentum orbits 
$\tau_R$ is slightly, but not crucially longer.}. Specifically, for a 
solar type star we obtain $\tau_R \sim $ 13~min, which is an order 
of magnitude less than the dynamic time scale $\tau_d \approx $ 3 hours.
The ratio of the two times indicates that the amount of energy exchanged may not be quite negligible,
but is small enough that it may be neglected in the first approximation.
Further justification for such an approximation comes from results of hydrodynamic evolution 
calculated by Kochanek (1994) and Laguna et al. (1993). Laguna et al 
noted that "the qualitative features of the debris - including its 
crescent-like shape - can be reproduced by neglecting hydrodynamic 
interactions and self-gravity of the star", since the formation of the 
crescent is due to "geodesic motion of the fluid elements of the 
star in a Schwarzschild space-time which includes relativistic-induced
precession of the orbit about the black hole".
This confirms previously mentioned findings by Luminet \& Marck (1985), that 
black hole's gravity dominates in close encounters. Therefore, we argue that 
by neglecting hydrodynamic effects, we obtain in close encounters approximately the
correct shape of stellar debris.

Hence, our numerical model starts with a spherically symmetric star 
of radius $R_*$ and mass $M_*$ consisting of $N$ equally massive 
constituents ($m_i=M_*/N$, $N\sim \rm{10^6}$) distributed randomly but 
in such a way that in the average their density 
distribution follows that of a star, which is approximated by the polytrope model with n=1.5.
All constituents of the star start with the (same) velocity, corresponding to 
parabolic velocity of the stellar center of mass, which is placed at a distance 
$\mathcal{R}_R$ from the black hole. Subsequently the positions of  
free falling stellar constituents are calculated at later discrete 
times ($\rm{t_i}$) according to general relativistic equations of motion.

\begin{figure*}
\centering
\includegraphics[width=19cm]{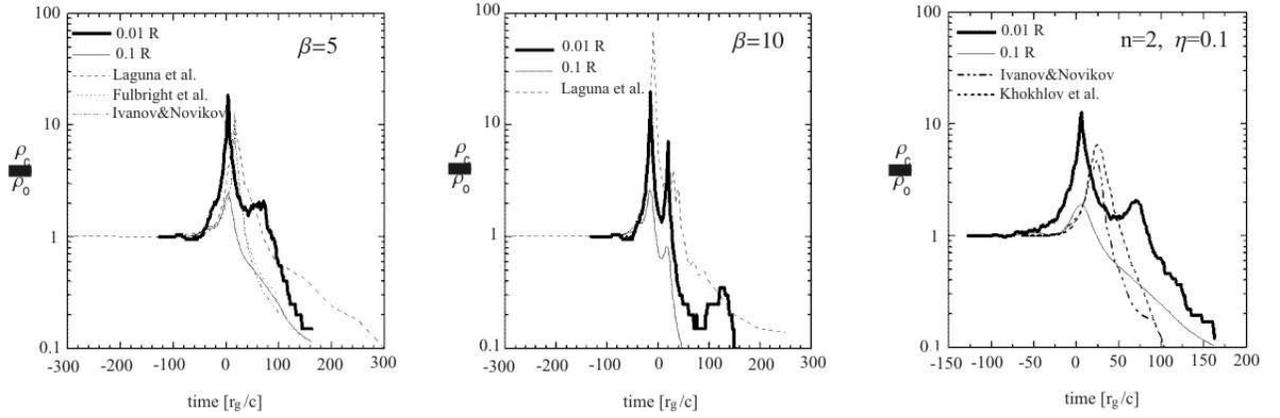}
\caption{The central density in the star as a function of time during close encounter, 
for polytrope n=1.5: $\beta$=5 ($\tilde{l}$=5), $\beta$=10 ($\tilde{l}$=4.08) and n=2, $\eta=0.1$ 
 ( $\eta = \bigl({{M_*}\over {m_{bh}}}\bigr)^{1/2}\, \bigl({{r_p}\over {R_*}}\bigr)^{3/2}$). 
Solid curves are from our simulations, dashed are
from Laguna et al. (1993), dotted from Fulbright et al (1995),
dot-dashed from Ivanov \& Novikov (2001) and short-dashed from Khokhlov et al. (1993).
Time is measured from the periastron passage. (Full resolution images available at www.fmf.uni-lj.si/\~~gomboc)}
\label{density}
\end{figure*}

To test the errors induced by these approximations, we tudy encounters of a 
M$_\odot$, R$_\odot$ star with a 10$^6$ M$_\odot$ black hole and compare our results on
central density in the star (average density inside 0.01 $R_*$) with those 
obtained by Laguna et al (1993), 
Fulbright et al. (1995), Khokhlov et al (1993b), Ivanov \& Novikov (2001).
Fig. \ref{density} shows the central density as a function of 
time with respect to the time of periastron passage as obtained by our model and by hydrodynamic 
simulations. The qualitative agreement between these results justifies the neglect of internal 
pressure in calculating the dynamics of disruption.
The major difference seems to be in the precise timing
of tidal compression: in our model the strongest compression occurs very close to periastron,
in agreement with the results of Luminet \& Marck (1985), 
while in most hydrodynamics simulations the central density peaks approximately 15-20 $r_g$/c 
after the periastron passage.

Our results on the shape of stellar debris during the close encounter
also agree with results of Laguna et al. (1993), although at later times
our crescent becomes considerably longer.

\subsection{On the luminosity of the star during the tidal disruption}

We consider the tidal disruption to be the phenomenon, where the work done on the star by tidal 
forces is comparable or greater than its initial internal energy. 
The tidal disruption is thus a violent nonstationary process that takes place in the vicinity of 
the black hole on a time scale  that is considerably shorter than the stellar dynamic time scale 
(measured in proper time of the falling star). As the star is deformed into a long thread, 
the giant tidal wave deposits great amounts of energy which soon pushes gases into an out moving 
shock wave more or less perpendicular to the threadlike axis of the star. Thus, during the disruption 
process several mechanisms play an important role: 
shocks, adiabatic expansion and cooling of disrupted material, possible explosions due to
tidal squeezings as predicted by Carter \& Luminet (1982, 1985), radiation driven expansion etc.
These effects have no doubt important influence on the cooling and luminosity
of the disrupted star, but we wish to stress, that as shown by Luminet \& Marck (1985)
gravity in general overwhelms other forces during the close encounter. So, since gravity of the black 
hole swings the star around on a time scale that is much shorter than any other time scale that may 
play a role, we believe, that as a first step to estimate the
luminosity variations of the tidally disrupted star, we may use a simple model, which must in the 
first place correctly take into account the effects of dominating strong gravitational field of the 
black hole.
As the disruption progresses and the hot stellar inner layers become exposed both by gravity and by 
shock waves, the luminosity is bound to rise 
due to higher effective temperature and due to higher effective area seen. The overall rise in 
luminosity depends
on other partially competing mechanisms involved: while the expansion and cooling would tend to reduce it,
it must nevertheless rise dramatically due to enormous work being done by tidal forces which 
drive shock heating and supernova-like explosions.
The precise role of these mechanisms and their influence on stellar structure and evolution need 
detailed analysis, but is beyond the scope of this paper.

\begin{figure*}
\centering
\centerline{\includegraphics[width=8cm]{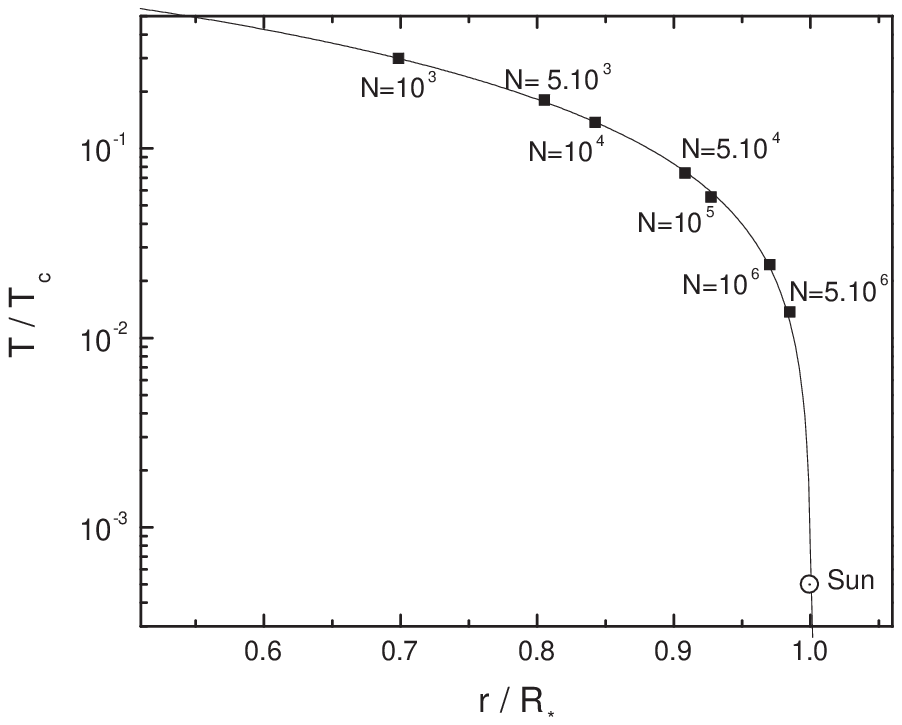} \includegraphics[width=8cm]{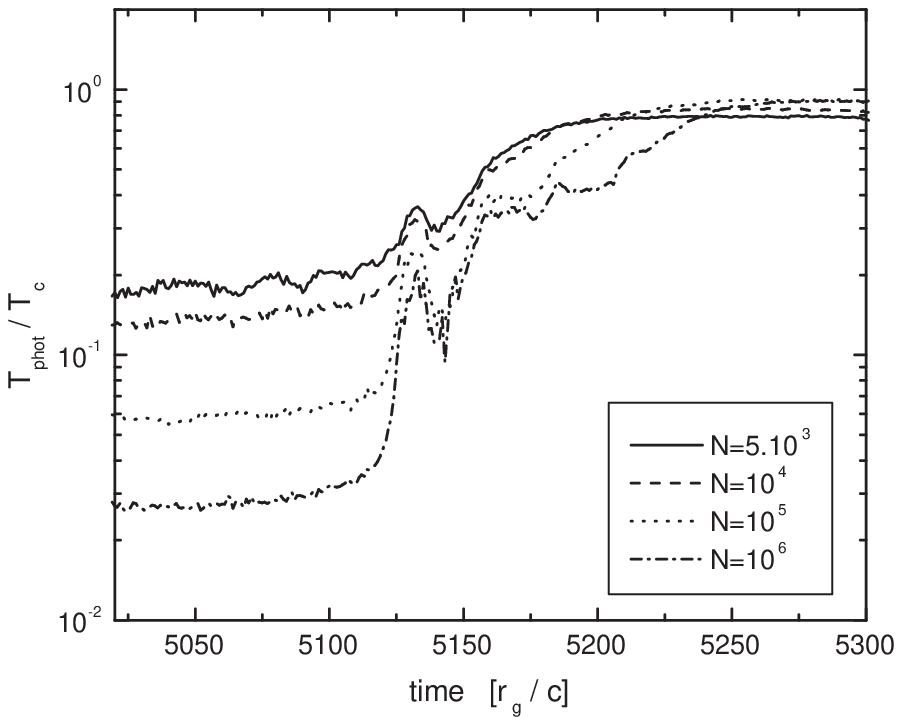}}
\caption{Left: The model atmosphere depth in the initial (spherically symmetric) star: 
polytrope temperature profile (line) and
"photospheric" temperatures and corresponding depths for different N in our model (symbols). 
Right: "Photospheric" temperature as a function of time during total tidal disruption given by our model for different N.}
\label{depth}
\end{figure*}

Here we wish to make a step towards the complete solution by including in full only the most 
important ingredient defining the shortest time scales: the effects
of black hole's gravity on the apparent variability of stellar luminosity. The standard stellar 
atmosphere model (Bowers and Deeming 1984, Carroll and Ostlie 1996, Swihart 1971) is not applicable in calculating the effective 
temperature of any surface element since, because of the highly dynamic structure, the fine details 
of atmospheric density, temperature and pressure profiles are not available, even more we can not 
predict in advance which part of the star is going at some future time belong to the atmosphere. 
So we are forced to apply a Monte Carlo model throughout the star by which the unperturbed star is 
modeled as a spherical cloud consiting of a large number ($N$) of identical constituents distributed 
randomly, but in such a way that their average density follows that of an $n=1.5$ polytropic model 
(cf. section 3.1). The constituents are optically thick and have an assigned temperature according to their position in the 
cloud, which again follows the temperature profile of the $n=1.5$ polytropic stellar model. 
The model "photospheric" temperature and model "luminosity" are calculated as the sum of spectral 
contributions from those cloud constituents that are seen by the observer, i.e. by those that are not 
obscured by constituents in overlaying layers. For the purpose of obscuration all the constituents 
are considered to have the same cross section $\sigma$, so that $\sigma=4 \pi R_*^2/N^\prime$, where the
parameter $N^\prime $ is the number of constituents belonging 
to "the atmosphere" of the star.
%is $N^\prime=4 \pi R_*^2/\sigma$. 
It is clear that, since for statistical reasons
$N^\prime $ must be at least a few ten, and $N$ is limited by the computer power to a few million, the ratio $N^\prime/N$ is 
much greater than the ratio $M_{atm}/M_*$ in a real star. One could argue that the atmosphere could be made less massive by representing it with a larger number of less massive constituents. However, in the case of total tidal disruption the interior is mixed into the atmosphere during the late stages of disruption and the so introduced uneven opacity of stellar constituents would further complicate the interpretation of results. Thus we can not afford to make models with 
sufficiently opaque atmospheres and, as a result, our initial model "photospheric" temperatures are too 
high. We note, however, that the model photospheric depth is a function of $N^\prime/N$, so by 
changing $N$, we probe the stellar atmosphere to different depths. In such a way an extrapolation to 
realistic opacities is possible. The consistency of such an extrapolation is checked on the initial 
spherically symmetric stellar model, where the Monte Carlo results can be directly compared with the 
theoretical atmospheric model. An example of such a comparison is shown in Fig.\ref{depth}. It is clear that 
the  depth of our model "photospheres" is some orders of magnitude too high, yet it is possible to 
extrapolate model "photospheres" to depths of realistic stellar atmospheres, since the temperature 
is a monotonic smooth function of depth. For evolved stages of tidal disruption there is no underlying
theoretical model, so that we rely on extrapolated results of the Monte Carlo model.

As the star moves along the orbit,
images of the star with respect to the far observer are
formed as follows: Photon trajectories and the time of flight between each
stellar constituent and the observer are calculated with technique described 
in \v Cade\v z et al. (2003), Gomboc (2001), Brajnik (1999) and \v Cade\v z \& Gomboc (1996). 
Only two trajectories connecting two space points are considered - the shortest one and the one 
passing the black hole on the other side, while those winding around the 
black hole by more than $\rm{2\pi}$ are neglected. (It has been shown before 
(\v Cade\v z \& Gomboc 1996), that light following  
trajectories with higher winding numbers contributes less and less to the apparent luminosity.) 
The beam contributions are sorted into pixels with an area corresponding to the size of $\sigma$, and tagged according to the arrival time. The intensity corresponding to a given pixel is then defined as the intensity corresponding to the ray with shortest travel time. Since light from deeper layers takes longer to reach the observer, this takes care of the obscuration of deep layers. %image frames according to the arrival time. Since photons from stellar interior need longer time to reach the outside observer than photons from stellar surface, we take care of obscuration of deep layers by keeping in each point of the star's image only the photon (beam) with the shortest travel time.
The intensity of a contributing beam is 
calculated assuming that the corresponding stellar constituent emits in its own rest frame 
as a black body at its temperature. The apparent luminosity and effective temperature 
of the star as a function of time (with respect to the chosen observer) are 
calculated and successive stellar images, formed in this way,
are pasted in a movie\footnote{Movies can be obtained at www.fmf.uni-lj.si/\~~gomboc}. 

We divide our model in three parts:
first we estimate the relativistic effects alone by simulating the luminosity variations
of an iso-thermal star (i.e. star with T(r,t)=const.). In the next step, we consider
the star with polytrope n=1.5 temperature profile and estimate the 
luminosity variations due to exposure of inner hot regions of the star. We first consider a
simple case, in which the temperature of all stellar constituents is constant in time
(no cooling or heating), and afterwards add a rough estimation of the effect of cooling of 
exposed stellar parts on the stellar luminosity.

\subsubsection{Effects of black hole's gravity}

\begin{figure*}
\centering
\includegraphics[width=17cm]{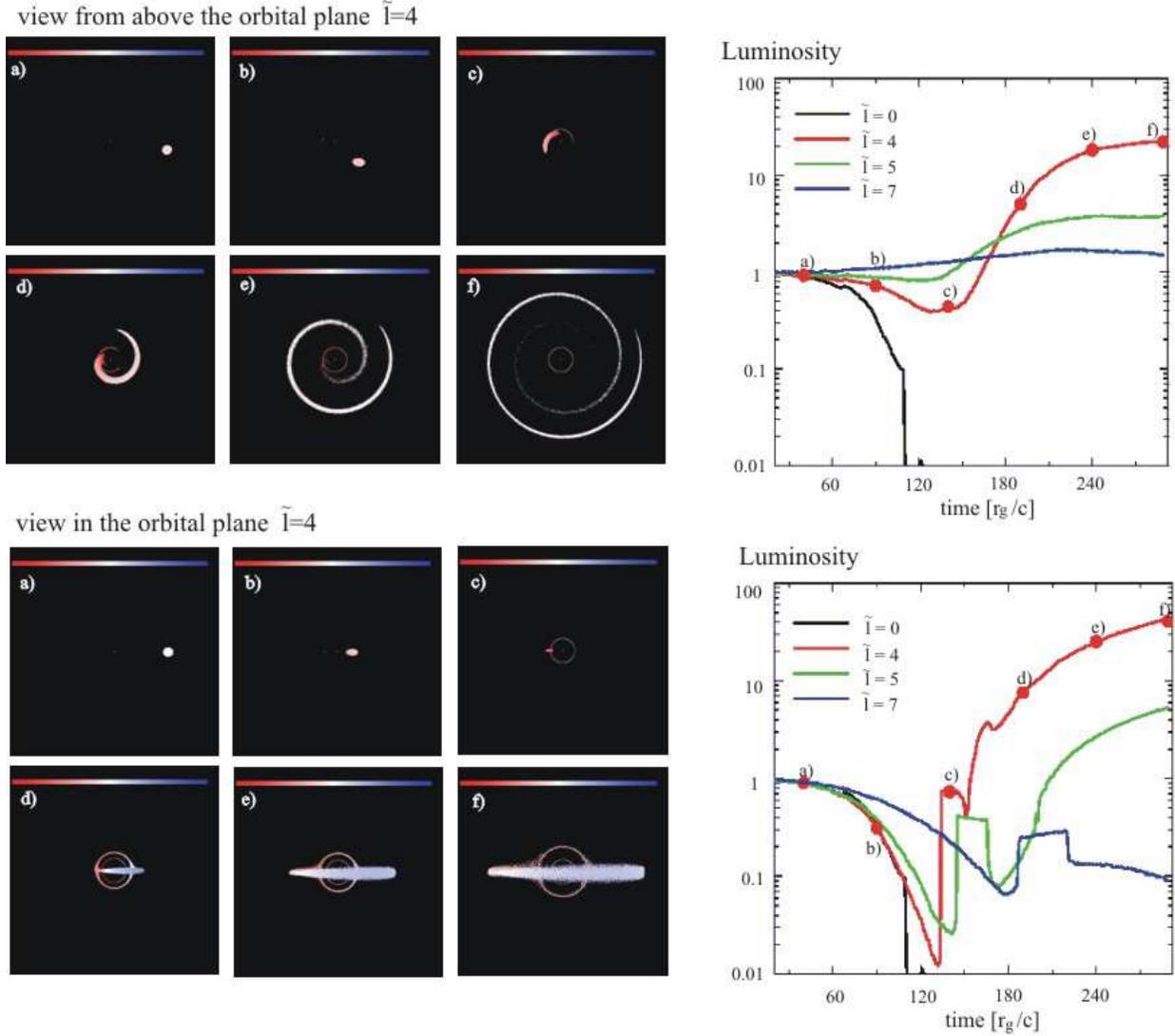}
\caption{Iso-thermal star with $R_*$=2$r_g$ during the encounter with critical $\tilde{l}$=4. 
Upper set is for 
the observer 
perpendicular to the orbital plane and the lower set is for the observer in the orbital plane.
Pictures show the stellar
appearance at time intervals of 50 $r_g$/c, with colour corresponding to the apparent temperature:
gravitational redshift close to the black hole and Doppler shift of receding material
stretch the observed frequency
of photons (and therefore the observed temperature of the stellar surface) towards
zero (red in colour code), while Doppler shift of approaching
material increases the observed temperature (blue colour, corresponds to the 
value of twice the temperature in the system comoving with the star). 
Graphs show the apparent
luminosity at different stages of the encounter 
(in units of the stellar luminosity before the encounter) and for different orbital angular 
momenta $\tilde{l}$=0, 4, 5, 7. (Full resolution images available at www.fmf.uni-lj.si/\~~gomboc)
}
\label{iso}
\end{figure*}

To isolate the effect of gravity, we first compute 
luminosity variations of an iso-thermal star. The ensuing luminosity variations can be ascribed to:  
Doppler boosting
and aberation of light, gravitational lensing and redshift (similar as for a 
point-like source in Sect. 2) and (in addition) the elongation of the star due to 
relativistic precession and due to tidal squeezing. Fig. \ref{iso} shows
the obtained luminosity variations as a function of time for encounters with $\tilde{l}$=0
(radial infall), $\tilde{l}$=4 (critical), $\tilde{l}$=5 ($r_p$=10 $r_g$) and $\tilde{l}$=7 
($r_p$=22.3 $r_g$)  
as seen perpendicular to and {\it in} the orbital plane.

Results show that the maximal rise in luminosity occurs in the case of the 
critical encounter ($\tilde{l}$=4), where the overall luminosity rise 
due to elongation of the star is
of about a factor of 20 (as seen by the observer perpendicular to the orbital plane, 
Fig. \ref{iso} above),
while gravitational lensing and Doppler boosting enhance it
up to about 40 times the initial luminosity (Fig. \ref{iso} below). 
Observers close to the orbital plane 
see most extreme variations: dimming of the receding star, its rebrightening as it
emerges from behind the black hole and variations on short timescales of 
about 10 $r_g$/c, which are due to lensing effects. 
Since the star and the black hole are comparable in size, the probability 
that they are aligned with respect to the observer, is high. When lensing 
takes place the relevant part of the star is imaged into an Einstein disk 
and the apparent luminosity increases manifold (Fig. \ref{iso}c).

\subsubsection{Constant temperature debris}
Next, we consider the star with n=1.5 polytrope temperature profile and we assume that
the temperature of stellar debris does not change with time. 
The model is obviously much too crude to rely upon its results regarding the spectral 
characteristics or even the absolute value of the emitted luminosity. The crude argument 
why this model may bear some resemblance to the true light curve is that shock wave released 
by the unbalance of gravity carries internal energy to the surface in such a way that the energy 
influx from the interior temporarily compensates the radiation loss.

\begin{figure*}
\centering
\includegraphics[width=17cm]{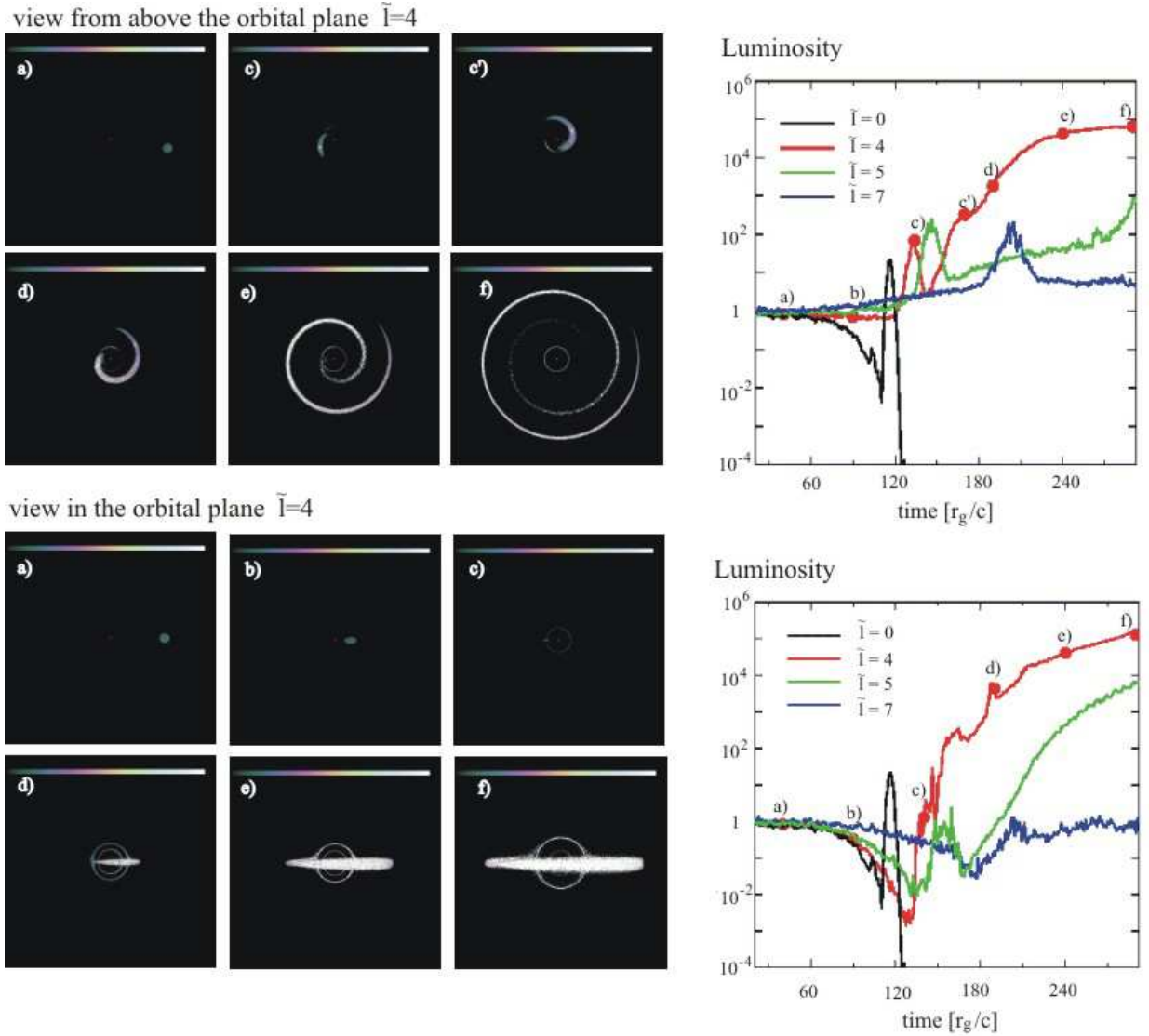}
\caption{Star with $R_*$=2$r_g$ and $\tilde{l}$=4 during the encounter assuming no temperature change 
of the debris.
Pictures show the stellar
appearance at time intervals of 50 $r_g$/c (except c and c', see graph) 
with colour corresponding to the apparent temperature
according to the colour code: blue - temperature zero, white - 0.5 T$_c$ or higher.
Inset graph shows the apparent
luminosity at different stages of the encounter (in units of the initial luminosity
far from the black hole) and for different orbital angular momenta $\tilde{l}$=0, 4, 5, 7.
Upper set is for the observer perpendicular to the orbital plane and the lower set is
for the observer in the orbital plane. (Full resolution images available at www.fmf.uni-lj.si/\~~gomboc)}
\label{konst}
\end{figure*}

The simulation shows that, as the inner hot layers 
of the star are exposed during the disruption, they contribute 
to the substantial rise in stellar luminosity, depending on the  
orbital angular momentum 
of the star (Fig. \ref{konst}). 
The star on a low angular momentum orbit is completely captured 
by the black hole and produces only a short ($\sim$ 1 - 10 $r_g$/c) flare before 
disappearing behind the horizon. On the other hand, the star with high 
angular momentum experiences only a slight distortion during the distant
flyby with a resulting temporary ($\sim$ 10 - 100 $r_g$/c) slight increase 
in luminosity. 

The most dramatic is the encounter of the star on the
critical angular momentum orbit ($\tilde{l}$=4), during which half of stellar constituents 
are swallowed by the black hole and the other half escapes. During this 
process, the star is totally tidally disrupted in such a way  that the 
higher angular momentum material rapidly lags behind the stellar debris 
with lower angular momentum, which produces a long thin spiral (Fig. \ref{konst}.).
Outer layers of the star are stripped off in a time of the order of 100 $r_g$/c, the depth to the hot 
inner core decreasing together with self gravity. In our crude model this is seen as decreasing 
optical thickness and the exposure of the 
hot inner core; the luminosity rises steeply.
The spectrum of the debris is dominated by the emission of the innermost 
exposed layers and as long as shock waves are building up, i.e. until cooling sets in, these lead 
to X-rays.

Some luminosity peaks arise from the effect of tidal compression in the 
direction perpendicular to the orbital plane of the star, which in our 
model for a short time exposes the interior of the star. Such 
peaks are evident in Fig. \ref{konst} above 
c and c', and these two compressions are in 
agreement with multiple tidal squeezings predicted by Luminet \& Marck (1985)
and confirmed by Laguna et al. (1993). In 
our model they produce luminosity peaks lasting about $\approx$ 5 $r_g$/c. 
As mentioned earlier, Carter \& Luminet (1982, 1985) predict, 
that thermonuclear explosion may occur at this moment.

The scale of the luminosity rise in Fig. \ref{konst} is rather uncertain due to neglect of hydrodynamic effects\footnote{For simplicity, we assume that all the tidal energy is transformed into the kinetic energy of the tidal wave; the portion of kinetic energy that may go into heat is neglected, therefore, we expect that the actual available luminous energy during such a tidal disruption may be higher than the one given by our models.} and also due to our poor atmosperic model (section 3.2.). For the critical tidal disruption of the Sun the extrapolation of our model would suggest the total luminosity to rise to about $10^{13}L_\odot$ (mostly in X-rays), which accentuates the extent of tidal disruption, but also sends a warning that by that time our constant internal energy model assumption ceases to be valid. As suggested in section 3.2., we calculated a range of models with $N$ between 10$^3$ and
$10^6$ and extrapolated the results to realistic atmosperic depths. These numerical results suggest that, at least for the critical disruption, the average temperature and size  of the final crescent to which the star is deformed is roughly
independent of $N$. Thus we tested the idea that
tidal disruption exposes or mixes up by shearing the envelope of the star to a certain depth
$R_c$, which we define as the depth in the undisturbed star, down to which the average $T^4$ is
equal to the average $T^4$ of the final crescent. In this way we estimate (independent of $N$) 
that for 
critical $\tilde{l}$=4 and $n=$1.5, $R_c$ is about 0.25 $R_*$, while for $n=5$ we get
$R_c=$0.1 $R_*$. For a close flyby with $\tilde{l}$=5, $R_c$ is about 0.7 $R_*$ and 0.5 $R_*$ for $n=$1.5 
and $n$=5 respectively.  We may also, as an example, estimate the luminosity of a Solar type star during the bright critical stage of total disruption on a 10$^6 M_{\odot}$ black hole as follows: the steep luminosty rise (c.f. Fig. \ref{konst}) has a time scale between 30$r_g$/c to 100$r_g$/c, which is about 2.5 to 8 minutes. Assuming that the initial thermal energy contained in the exposed layers ($\sim$10$^{48}$ erg) is radiated away on this time scale, the critical luminosity would be of the order 5 to 15$\times 10^{11}L_\odot$.

After the debris is spread and starts moving away from the black hole, the physics of tidal disruption 
is no longer dominated by black hole's gravity. The physical conditions in stellar debris, the physics 
of radiation processes, hydrodynamics etc. take over and the ensuing processes go beyond the 
simulation presented here.
\subsubsection{Cooling of stellar debris}
In general, the temperature inside the star may change due to various mechanisms, already mentioned.
To get an idea of how they might affect the light curve, we again model 
the cooling in two very approximative ways: 

(a) exponential cooling of exposed stellar layers with different characteristic times: 
$\tau$=1 $r_g$/c and 10 $r_g$/c,

(b) cooling of exposed stellar layers as due to their own black body radiation in the 4$\pi$ solid angle. 

\begin{figure*}
\centering
\includegraphics[width=14cm]{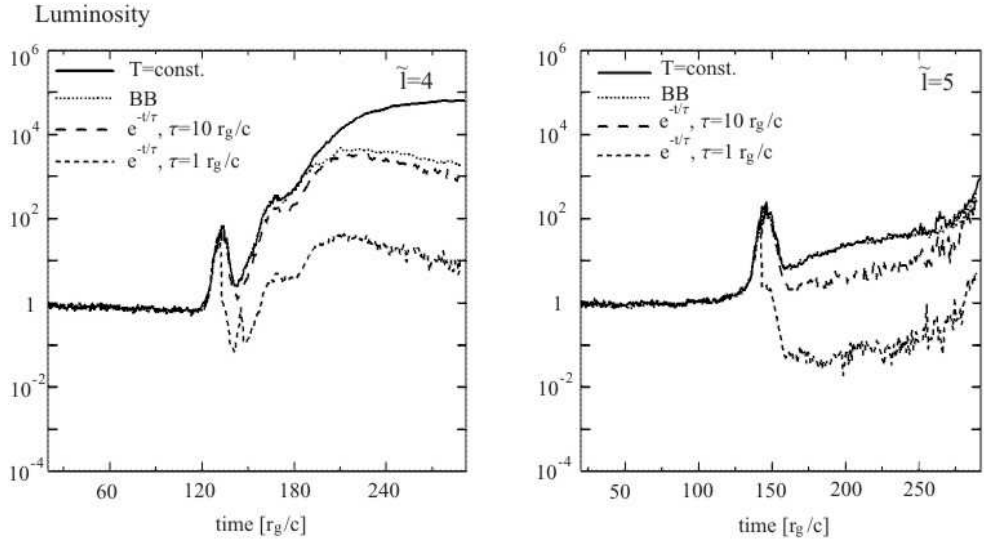}
\caption{Effect of cooling of debris on the luminosity variations: solid line: no cooling, dotted: cooling by
black body radiation, dashed lines: exponential cooling with decaying time 10 $r_g$/c (long dash) 
and 1 $r_g$/c (short dash).
Results are for the star on $\tilde{l}$=4 (left) and $\tilde{l}$=5 orbit (right) as observed 
perpendicular to the orbital plane. (Full resolution images available at www.fmf.uni-lj.si/\~~gomboc)}
\label{cool}
\end{figure*}

Results presented in Fig. \ref{cool} show, that if the cooling were very efficient -
with timescales of 1 $r_g$/c, the luminosity rise would be quite short and modest.

\section{Conclusion}
Stellar encounter with a massive black hole can be a very energetic event, 
with energy released and luminosity variations depending primarily on the relative size of 
the star compared to the black hole. We note that the tidal interaction energy may rise to 
as high as 10\% of the total mass-energy of the captured star, which is available 
when the star is comparable in size 
to the size of the black hole. This size ratio is also critical as to the nature of the disruption. 

In this work we focused on gravitational phenomena and showed that: 
\begin{itemize}
\item{a)}
A critical capture of a "pointlike star" is characterized by a series of quasiperiodic apparent 
luminosity peaks with the quasiperiod 50 $r_g$/c for a Schwarzschild black hole and 13 and 
80 $r_g$/c for an 
extreme Kerr co- and counter-rotating case respectively (Fig. \ref{f1}). This translates into
6.9 hours $\times \rm{{m_{bh}}\over {10^8 M_\odot}}$, 1.8 hours $\times \rm{{m_{bh}}\over 
{10^8 M_\odot}}$ and 11.1 hours $\times \rm{{m_{bh}}\over {10^8 M_\odot}}$,
respectively.
If a "pointlike star" would be a planet falling to a $3.6\times 10^6 M_\odot$ black hole
in Galactic centre respective quasiperiods would be 15 minutes, 3.9 minutes and 24 minutes.
\item{b)}
The sharpness, the amplitude of quasiperiodic peaks and the amplitude of the Doppler factor 
is more pronounced for observers in the orbital plane as compared to those perpendicular to this plane. The highest value for the Doppler factor is 1.8 for the Schwarzschild and 2.2 for the extreme Kerr black hole. 
\item{c)} 
The number of quasiperiodic peaks ($N_p$) depends on the closeness of the orbital angular momentum 
($\tilde{l}$) to the critical value $\tilde{l}$=4 and can be 
approximated as $N_p=0.5-0.5 Log(4-\tilde{l})$.
\item{d)}
An extended star may be approximated as a collection of point particles when heading toward the complete tidal disruption. The shape and the density of the debris calculated in this approximation compares well with more sophisticated hydrodynamic calculations (cf. Section 3). 
\item{e)} 
Model light curves for critical tidal disruption of a star of the same size as that 
of the black hole (Fig. \ref{iso}, \ref{konst}, \ref{cool}) calculated for different heuristic 
models show similar temporal characteristics which display very rapid (on time scale of order 10 $r_g$/c)
luminosity variations by a few or even many orders of magnitude, while the quasiperiodicity is no 
longer pronounced in such a process. 
Light curves describing a critical capture are very rough 
and can not be momentarily calibrated in flux. They are presented as they produce the extremely short time scale phenomena characteristic of the strength of black hole's gravitational field, which will persist in the future more elaborate models of tidal disruption.  
\end{itemize}

\begin{appendix}
\section{The virial theorem and tidal energy}
In order to estimate the amount of heat and kinetic energy deposited to the star by the tidal wave, it is useful to follow the steps of the  derivation of the virial theorem. Consider the some $10^{60}$ nuclei and electrons making up the star as representative point particles making up the ideal gas of the star. Each of the particles with mass $m_i$ ($i~=~1\dots 10^{60}$) moves according to Newton's law (We will folow the more transparent classical derivation, which is sufficient for order of magnitude arguments.):
\begin{equation}
m_i\ddot {\vec r_i}~=~\sum_{j\neq i}\vec F_{ij}^c + \sum_{j\neq i}G{m_i m_j \over \vert \vec r_j - \vec r_i\vert ^3 }(\vec r_j - \vec r_i)-G {{m_{bh}m_i}\over r_i^3}\vec r_i
\label{Newton}
\end{equation}
The black hole has been placed at the origin from where the position vectors $\vec r_i$ are reconed. $\vec F_{ij}^c$ models the force taking place during particle collisions. It obeys (the strong version of) the third Newton's law, and since in the ideal gas approximation collisional forces act only at a "point", the energy connected with the potential of these forces can be neglected. The second term on the right describes the gravitational interaction among the constituents of the star and the last term represents the gravitational force of the black hole. It is convenient to define the center of mass position vector $\vec R~=~ \Biggl (\sum_i m_i \vec r_i\Biggr )/M_*$, so that $\vec r_i~=~\vec R+\vec r_i^\prime$ and $\sum_i m_i \vec r_i^\prime~=~0$. Summing equations \ref{Newton} over all $i$, one obtains the center of mass equation of motion in the form:
\begin{equation}
M_* \ddot{\vec R}~=~-G {m_{bh}M_* \over R^3}\vec R - 5 G m_{bh}{\vec R.{\bf Q}.\vec R \over R^7}\vec R + 2 G m_{bh} {{\bf Q}.\vec R \over R^5}+{\cal O}(1/R^5),
\label{CM}
\end{equation}
where ${\bf Q}$ is the quadrupole moment tensor of the mass distribution with respect to the center of mass of the star defined in the usual way as:
\begin{equation}
{\bf Q}~=~{1\over 2}\sum_i m_i(3 \vec r_i \vec r_i -{\bf I}r_i^2).
\label{quadru}
\end{equation}
Terms of ${\cal O}(1/R^5)$ and higher will henceforth be neglected.
If the star is deformed in a prolate ellipsoid with the long axis in the direction $\hat n$, ${\bf Q}$ can be written in the form 
\begin{equation}
{\bf Q}~=~3 q \hat n \hat n-q{\bf I}~~~~,
\label{kvad}
\end{equation}
with $q$ beeing positive and proportional to the eccentricity of the ellipsoid. 
Here $\vec r_i \vec r_i$
stands for the diadic product of the respective vectors and $\bf I$ is the identity matrix.

The angular momentum of the star ($\vec l$), which is a conserved quantity, can be split into the orbital ($\vec l_o = M \vec R\times \dot{\vec R}$) and spin part ($\vec l_s = \sum_i m_i \vec r_i^\prime \times \dot{\vec r_i^\prime}$). The time derivative of the orbital part follows from eq.\ref{CM} and when \ref{kvad} applies, it can be written as:
\begin{equation}
\dot{\vec l_o}~=~6 G{ m_{bh}q\over R^5}(\vec R \times \hat n)(\vec R.\hat n)
\label{angmom}
\end{equation}

The sum of scalar products of equations \ref{Newton} by $\dot{\vec r_i}$ gives the energy conservation law. We split the kinetic energy of the star into the center of mass part ${1 \over 2}M_*\dot{\vec R}^2$ and the internal kinetic energy part\footnote{Note that $W_{int}$ comprises both the kinetic energy of thermal motion and the kinetic energy of bulk motion in the tidal wave.} $W_{int}~=~\sum_i {1\over 2}m_i \dot{\vec r_i^\prime}^2$. Using eq. \ref{CM} and neglecting the collisional interaction energy, we obtain the conserved energy $E$ in the following form:
\begin{equation}
E~=~{1\over 2}M_* \dot{\vec R}^2-G{m_{bh}M_*\over R}-G{m_{bh}\vec R.{\bf Q}.\vec R\over R^5}+W_{int}+W_G, 
\label{energ}
\end{equation}
where $W_G$ is the self gravitational energy of the star ($W_G=-{1\over 2}\sum_i\sum_{j\neq i}G{m_i m_j\over \vert {\vec r_j}^\prime -{\vec r_i}^\prime \vert}$)

Finally, we obtain the equivalent of the virial theorem by we multiplying eqs.\ref{Newton} by $\vec r_i^\prime$ and summing over all $i$. The result can be rearanged into the transparent form : 
\begin{equation}
W_{int}+{1\over 2}W_G~=~-G {m_{bh}\vec R.{\bf Q}.\vec R\over R^5}+{1\over 4}\ddot J~~~~,
\label{fvir}
\end{equation}
where $J~=~\sum_i m_i {r_i^\prime}^2$. 
For a star in hidrostatic equilibrium, the right-hand side vanishes and the total energy of the star $W_{tot}~=~W_{int}+W_G~=~-W_{int}$. If the star is not in hidrostatic equilibrium, the right hand side of eq.\ref{fvir} can be considered as the energy imbalance - if it is more than $W_{int}$, it is sufficient to completely disrupt the star on a time scale $\tau_d$. An exact evaluation of this energy imbalance is beyond reach in this simple analysis, however, a simplified model offers some clues. 

Consider an idealized case of an "incompressible star" flying about a massive black hole. From the point of view of the star, gravity is exerting a tidal force squeezing it in the plane defined by the temporary radius vector and the orbital angular momentum and elongating it perpendicular to this plane. The tidal force acts to accelerate the surface of the star with respect to the center of mass, but it must also act against rising pressure and internal gravity. Thus, roughly speaking, the tidal force does work in pumping kinetic energy into the tidal wave, but also in loading the gravitational potential energy which acts as the spring energy driving oscillation modes of the star. Consider small tidal distortions. In this case quadrupole deformations are dominant, so that the deformation field ($\vec U$) of the incompressible star can be described as a linear combination of 5 degenerate quadrupole modes:
\begin{equation}
\vec U ~=~\sum_{k=1}^5 a_k \vec {\cal q}_k~~~~.
\label{modedec}
\end{equation}
Here $\vec {\cal q}_k$ are modal base vector fields that can be expressed as gradients of quadratic polynomials in coordinates $x^\prime$, $y^\prime$, $z^\prime$,  obtained by multiplying spherical functions $Y_{2m}(\theta^\prime , \phi^\prime)$ by ${r^\prime}^2$ and identifying $x^\prime=r^\prime \sin \theta^\prime \cos \phi ^\prime$ etc, and $a_k(t)$ are modal amplitudes. In the coordinate system  where the 
$z^\prime$ axis is normal to the orbital plane and $x^\prime$ points from the periastron to the black hole, only three amplitudes are excited and the corresponding modal base fields are:
\begin{equation}
\vec {\cal q}_5= -\sqrt{5\over 4 \pi}\pmatrix{ -x,-y,2 z} \\~~~~~  
\vec {\cal q}_1= -\sqrt{15\over 4 \pi}\pmatrix{ x,-y,0}   \\~~~~~  
\vec {\cal q}_2= -\sqrt{15\over 4 \pi}\pmatrix{ y,x, 0}
\label{modes}
\end{equation}
These deformations lead to the following quadrupole moments:
\begin{equation}
{\bf Q}~=~{1\over 4 \pi}\pmatrix { {{a_1}}^2 + {{a_2}}^2 - 
    2\,{\sqrt{3}}\,{a_1}\,{a_5} - {{a_5}}^2,
   -2\,{\sqrt{3}}\,{a_2}\,{a_5},0\cr 
   -2\,{\sqrt{3}}\,{a_2}\,{a_5},
   {{a_1}}^2 + {{a_2}}^2 + 
    2\,{\sqrt{3}}\,{a_1}\,{a_5} - {{a_5}}^2,0\cr 
   0,0,-2\,\left( {{a_1}}^2 + {{a_2}}^2 - {{a_5}}^2
      \right) }
\label{Qji}
\end{equation}
As long as tidal modes can be considered roughly independent, their dynamics can be derived from the Lagrange function $L~=~T-U$ with the kinetic energy ($T$):
\begin{equation}
T~=~\sum_{k=1}^5 \sum_{l=1}^5 \int \rho \dot {a_k} \dot {a_l}\vec {\cal q}_k \vec {\cal q}_l dV^\prime~=~{3\over 4 \pi}M_* R_*^2 \sum_{i=1}^5 \dot{a_i}^2
\end{equation}
and the potential energy ($U$ - the deviation od self gravity from the equilibrium value in undeformed state):
\begin{equation}
U~=~{3\over 4 \pi}M_* R_*^2 \sum_{i=1}^5 \omega_q^2{a_i}^2~~~,
\end{equation}
where $\omega_q$ is the resonant frequency of quadrupole modes. For a star consisting of a self gravitating incompressible fluid we obtain 
\begin{equation}
\omega_q^2={64\over 5}GM_*/R_*^3~~~
\label{kvadruf}
\end{equation}
Generalized forces exciting these  modes are (Goldstein 1981):
\begin{equation}
F_k~=~\int \rho \vec {\cal q}_k G{m_{bh}\over R^3}.({\bf I}-3 {\vec R\over R} {\vec R \over R}).\vec {r^\prime} dV^\prime~~~,
\label{qji}
\end{equation}
Let us calculate these forces in the specific case when one can assume that $\vec R(t)$ represents a parabolic orbit.We express the compontents of $\vec R$ as 
\begin{equation}
\underline R(t)~=~R(t)\lbrace \cos \psi(t),\sin \psi(t),0\rbrace ~~~~~,
\end{equation}
where 
\begin{equation}
R(t)~=~ r_p/\sin^2{1\over 2}\psi(t)
\end{equation}
and $\psi(t)$ is the true anomaly obeying the Kepler equation: 
\begin{equation}
\sqrt{G m_{bh}\over 2 r_p^3}t~=~-\cot {1\over 2}\psi(1+{1\over 3}\cot^2{1\over 2}\psi)~~~~.
\end{equation}
With this, and using \ref{modes}, the integrals in \ref{qji} can be evaluated to obtain the nonvanishing generalized forces:
\begin{equation}
\pmatrix{{F_1(t)}\cr {F_2(t)}\cr {F_5(t)}}~=~-3 G {m_{bh}\over 16 r_p^3}\sqrt{3 \over 5\pi }M_*R_*^2\sin^6{1\over 2}\psi \pmatrix{\cos 2\psi \cr \sin 2 \psi \cr -1/\sqrt 3}
\end{equation}
Finally we write down the Euler-Lagrange equations of motion (${d\over dt}{\partial L\over \partial \dot{a_k}}-{\partial L\over \partial a_k}=Q_k$) for modal amplitudes. After introducing the characteristic time $t_f=\sqrt{2 r_p^3\over Gm_{bh}}$ and the dimensionless time $\tau=t/t_f$, they can be cast into the dimensionless form:
\begin{equation}
{d^2a_i\over d\tau^2}+(\omega_q t_f)^2 a_i~=~f_i(\tau)~~~,
\label{eqmotion}
\end{equation}
where the dimensionless forces $f_i(\tau)$ are functions of $\psi(\tau)$ only:
\begin{equation}
\pmatrix{f_1\cr f_2 \cr f_5}~=~-{1\over 4}\sqrt{3 \pi\over 5}\sin^6 {1\over 2}\psi\pmatrix{\cos 2\psi \cr \sin 2 \psi \cr -1/\sqrt 3}
\end{equation}
Thus, the only trace of parameters of the tidaly interacting system is left in the factor $\omega_q t_f$, which is $2\pi$ times the ratio of the characteristic fly-by time about the black hole and the period of quadrupole modes. It is useful to note that, using eq.\ref{Roche} and \ref{kvadruf}, this product can be written as 
\begin{equation}
\omega_q t_f~=~8\sqrt{2/5}(r_p/r_R)^{3\over 2}~=~8\sqrt{2/5}(1/\beta)^{3\over 2},
\label{omegabeta}
\end{equation}
i.e., it is inversely proportional to the power of Roche radius penetration depth. In the case of a distant fly-by $\omega_q t_f\gg 1 $, so it follows from eq.(\ref{eqmotion}) that $a_i=f_i(\omega_q t_f)^{-2}~\propto ~{m_{bh}\over \rho_* r_p^3}$, which is the familiar result often used with Earth tides. Note, however, that for deep penetrations of the Roche radius $\omega_q t_f\leq 1 $, and thus the (dimensionless) generalized forces $f_i(t)$ become large at frequencies that are resonant with $\omega_q$. 

We calculate the total work done by tidal forces on the system of normal modes during the whole fly-by process by noting that it can formally be expressed as the change of the Hamiltonian $H(t)~=~T+U$ during the process (neglecting damping of normal modes). Initially the quadrupole system starts in the undisrupted state with $H(t\rightarrow -\infty)=0$, and ends in a state of excited quadrupole modes with $W_{tid}=H(t\rightarrow \infty)~$\footnote{Assuming the tidal kick did not break up the star by imparting higher than escape velocity to surface layers.}
(i.e. for $t\gg t_f$):
\begin{equation}
W_{tide}~=~{3\over 4 \pi}M_* R_*^2\sum_{i=1}^5\lim_{t\rightarrow \infty}({\dot a_i}^2+\omega_q^2 a_i^2)~=~\sum_{i=1}^5 \int_{-\infty}^{\infty}F_i(t){\dot a_i}dt
\end{equation}
Solving equations \ref{eqmotion} with the retarded Green's function, this can be written in the form:
\begin{equation}
W_{tide}~=~{3\over 4} G{m_{bh}M_*R_*^2\over r_p^3}\sum_{i=1}^5 \vert  \hat {f_i}(\omega_q t_f) \vert^2
\label{tidalref}
\end{equation}
where 
\begin{equation}
\hat {f_i}(\Omega)~=~{1\over \sqrt {2\pi}}\int_{-\infty}^\infty f(\tau) e^{i\Omega \tau}d\tau
\end{equation}

\begin{figure*}
\centering
\includegraphics[width=17cm]{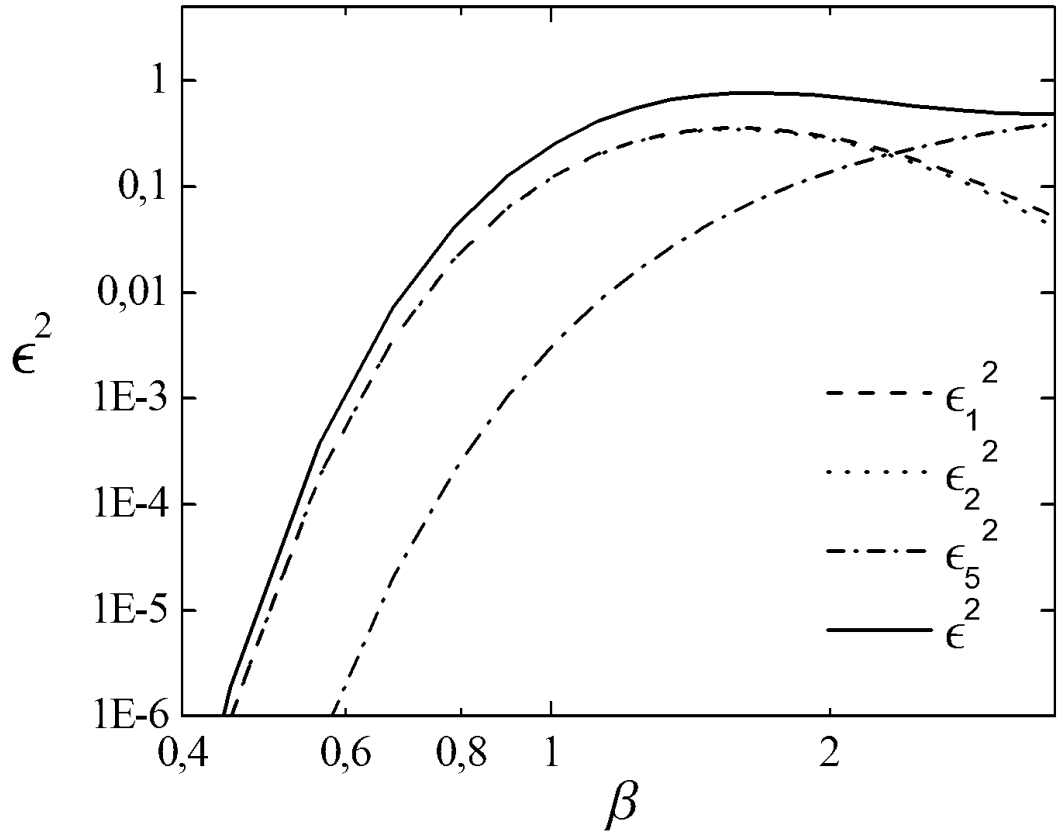}
\caption{The effective eccentricity $\varepsilon^2$ as a function of Roche penetration parameter. The lower three curves represent contributions due to
the three excited modes (1,2,5).}
\label{f20}
\end{figure*}

We note that $W_{tide}$ can be written in the form $G m_{bh}\tilde q/r_p^3$, where 
$\tilde q~=~M_* R_*^2 \varepsilon^2$ and according to (\ref{tidalref}) 
\begin{equation}
\varepsilon^2= {3\over 4}~\sum_{i=1}^5 \vert  \hat {f_i}(\omega_q t_f) \vert^2
\label{epsi}
\end{equation}
can be thought of as an effective eccentricity of the star at the periastron.
Fig. \ref{f20} shows that $\varepsilon$ can reach values of order $1$ if a fly-by is comparable to the dynamic time-scale of the star. Note however that for deep Roche radius penetrations our first order perturbation model no longer applies; closer analysis shows that the model is aplicable for $\omega_q t_f > 1$ i.e.  for $\beta \lesssim 3$ (eq. \ref{omegabeta})\footnote{We note that for $1 \lesssim \beta \lesssim 3$ the tidal energy is proportional to $\beta^2$ since $\varepsilon ^2 \propto 1/\beta$. This is in agreement with result of Lacy et al 1982 and Carter \& Luminet 1983. 
}.

Now we are in the position to estimate the high value of the right hand side of eq.\ref{fvir} for this simple parabolic infall of an incompressible star. The left hand side starts at zero, when the star is still far from the black hole. As time goes on, the internal kinetic and potential energy change as the energy of tidal modes, so that the left hand side is greatest when all the tidal energy is in the kinetic energy of the wave. Thus, the maximum value, which is also the maximum value of the right hand side equals $W_{tide}$.

Even if the above analysis is valid, strictly speaking, for an incompressible star and in the approximation of independent (small amplitude) tidal modes, it does suggest the qualitative conclusion that the tidal interaction depends crucially on the ratio period of the fundamental mode versus typical fly-by time ($\omega_q t_f$) and does become resonant if the fly-by time is less than the period of the fundamental mode. The energy deposited into the star by the tidal interaction can be of the order 
$G{m_{bh}M_* R_*^2\over r_p^3}~=~M_* c^2 r_g R_*^2/r_p^3$,
which may surpass the absolute value of the internal gravitational energy of the star by many orders of magnitude if $r_p$, $R_*$ and $r_g$ happen to be of the same order. 
\end{appendix}

\begin{acknowledgements}
We thank the anonymous
referee for constructive criticism which helped us improve the text.
We acknowledge the financial support of the Slovenian Ministry of Science, Education and Sport.
AG also acknowledges the receipt of the Marie Curie Fellowship from the European Commission.
\end{acknowledgements}

\clearpage

\clearpage

\clearpage

\clearpage

\clearpage

\clearpage

\end{document}